\documentclass[a4paper]{jpconf}

\usepackage{ amssymb }
\usepackage{ amsmath }
\usepackage{ relsize }
\usepackage{ hyperref }
\usepackage{ graphicx }
\usepackage{ xspace }

\newcommand{\PbPb}{\mbox{Pb+Pb}}
\newcommand{\pp}{{\it pp}}
\newcommand{\pPb}{\mbox{$p$Pb}}
\newcommand{\RpPb}{\mbox{$R_{p\mathrm{Pb}}$}}

\newcommand{\meansumet}{\mbox{$\langle \Sigma E_{\mathrm{T}}\rangle$}}
\newcommand{\ptavg}{\mbox{$p_\mathrm{T}^\mathrm{avg}$}}
\renewcommand{\pt}{\mbox{$p_\mathrm{T}$}}
\newcommand{\Et}{\mbox{$E_\mathrm{T}$}}
\newcommand{\etaavg}{\mbox{$\eta^\mathrm{avg}$}}
\newcommand{\aj}{\mbox{$A_{\mathrm{J}}$}}
\newcommand{\meanaj}{\mbox{$\langle A_{\mathrm{J}}\rangle$}}
\newcommand{\RdR}{\mbox{$R_{\Delta R}$}}
\newcommand{\rhoRdR}{\mbox{$\rho_{R_{\Delta R}}$}}

\newcommand{\ANpart}{\mbox{$\langle N_{\mathrm{part}}\rangle$}}
\newcommand{\ANcoll}{\mbox{$\langle N_{\mathrm{coll}}\rangle$}}
\newcommand{\Ncoll}{\mbox{$N_{\mathrm{coll}}$}}

\newcommand{\sqrtsnn}{\mbox{$\sqrt{s_{\mathrm{NN}}}$}}

\newcommand{\Raa}{\mbox{$R_{\rm AA}$}}

\newcommand{\sumet}{\mbox{$\Sigma E_{\mathrm{T}}$}}

\begin{document}

\title{Overview of new results from ATLAS heavy ion physics program}

\author{ Martin Spousta \\ on behalf of the ATLAS Collaboration  }

\address{Charles University in Prague, Institute of Particle and Nuclear Physics, \\ V Holesovickach 2, 180 00 Prague 8, Czech republic}

\ead{martin.spousta@cern.ch}

  \begin{abstract}
  In this short paper we provide an overview of new results from the ATLAS physics program at the LHC as of spring 2015. We separately summarize the results from \pPb\ 
collisions and \PbPb\ collisions along with some of their interpretations.
  \end{abstract}

\section{Introduction}
 
  In this short paper we provide an overview of new results from the ATLAS physics program at the LHC \cite{Aad:2008zzm}. 
  In fall 2010 and fall 2011, ATLAS has collected 7~$\mu$b$^{-1}$ and 0.14~nb$^{-1}$, respectively, of lead-lead (\PbPb ) collisions at \sqrtsnn\ = 2.76~TeV used for 
physics analyses. In fall 2012 a pilot proton-lead (\pPb ) run was realized delivering approximately 1 $\mu$b$^{-1}$ of collisions at \sqrtsnn = 5.02~TeV. The full 
\pPb\ run at the same center of mass energy followed in the winter 2013 delivering approximately 28 nb$^{-1}$. Some of the measurements done with \PbPb\ collisions use 
as a reference proton-proton (\pp ) collisions delivered at the same center-of-mass energy, that is $\sqrt{s} = 2.76$~TeV which were provided by the LHC in the year 2013. These 
\pp\ collisions together with \pp\ collisions at 7 or 8~TeV are used also to construct a reference at 5.02 TeV which is used in some of the \pPb\ measurements.
  The results from \pPb\ collisions are summarized in Sec.~\ref{sec:ppb}, the results from \PbPb\ collisions are summarized in Sec.~\ref{sec:pbpb}.

\section{Physics of \pPb\ collisions}
\label{sec:ppb}

  Prior to the first LHC runs of \pPb\ collisions, the main motivation for running these collisions was to improve the understanding of the initial state effects of nuclear 
collisions, namely to quantify the modifications of parton distribution functions in the nuclear environment. The studies done with \pPb\ data were primarily intended
to be a reference for studies of processes in \PbPb\ collisions, where a deconfined matter is produced. However, the \pPb\ collisions have brought several 
unexpected observations, such as the double-ridge structure in long range pseudorapidity correlations, significant centrality and rapidity dependence of the inclusive jet 
production, and an enhancement of charged particle yields.
  In this section we summarize the recent studies in \pPb\ collisions provided by ATLAS starting from measurements of jet yields, going through measurements of correlations 
between soft and hard processes, $Z$~boson, $J/\psi$ and $\psi(2)$ production, jet fragmentation, and closing by studies of the azimuthal correlations in \pPb\ 
collisions.

  Measurement of the inclusive jet production in \pPb\ \cite{Aad:2014cpa} was expected to provide a valuable benchmark for the jet quenching measured in \PbPb\ 
collisions. Indeed, a good correspondence of the jet spectra measured inclusively in centrality with the pQCD prediction employing the EPS09 parameterization of nuclear 
parton distribution functions (nPDF) \cite{Eskola:2009uj} was seen, which confirms that the jet suppression seen in \PbPb\ collisions is due to final state effects. The 
nuclear modification factor of inclusive jets, \RpPb , exhibits only little (if any) deviation from unity. On the contrary, the ratios of inclusive jet spectra from 
different centrality selections show a strong modification of jet production at all \pt\ at forward rapidities and for large \pt\ at mid-rapidity, which manifests as a 
suppression of the jet yield in central events and an enhancement in peripheral events. These effects imply that the factorization between hard and soft processes is 
violated at an unexpected level in proton-nucleus collisions. Furthermore, the modifications at forward rapidities were found to be a function of the total jet energy 
only, implying that the violations might have a simple dependence on the hard parton-parton kinematics.

  To improve the understanding of soft-hard correlations ATLAS measured the relationship between jet production and the underlying event in a pseudorapidity separated 
region in 2.76 TeV \pp\ collisions \cite{ATLAS:2015hwg}. In that study, the underlying event was characterized through measurements of the average sum of the transverse energy at large 
negative pseudorapidity, \meansumet , which were reported as a function of hard scattering kinematic variables. The hard scattering was characterized by the average 
transverse momentum, \ptavg , and pseudorapidity, \etaavg , of the two highest transverse momentum jets in the event. It was found that the \meansumet\ is anticorrelated 
with the dijet \ptavg , decreasing by 25\% as \ptavg\ varies from 50 to 500 GeV.
  This general trend is reproduced by leading-order Monte Carlo (MC) generators.
  These anticorrelations measured in \pp\ collisions provide a useful context 
for understanding the \pPb\ results, since they indicate a nontrivial correlation between hard scattering kinematics and \sumet\ production.

  Further insight was gained by estimating, from the dijet kinematics on an event-by-event basis, the scaled longitudinal momenta of the hard scattered partons in protons. This was done separately in the 
projectile and target beam-protons defined as moving to positive and negative rapidity, respectively. Transverse energy production at large negative pseudorapidity was 
observed to be linearly dependent on the longitudinal momentum fraction in the target proton, $x_\mathrm{targ}$, and only weakly with that in the 
projectile proton, $x_\mathrm{proj}$.
  This shows that the average level of transverse energy production is sensitive predominantly to the Bjorken-$x$ of the parton originating in the beam-proton which is 
headed towards the energy-measuring region ($x_\mathrm{targ}$), and is mostly insensitive to $x$ in the other proton ($x_\mathrm{proj}$). 
  These results provide counter-evidence to claims that the observed centrality-dependence of the jet rate in
\pPb\ collisions simply arises from the suppression of transverse energy production at negative rapidity
in the hard-scattered nucleon-nucleon sub-collision (being e.g. a consequence of an energy loss conservation \cite{Armesto:2015kwa}).
  In the \pPb\ data, the deviations from the expected centrality
dependence are observed to depend only on, and increase with, $x$ in the proton, that is with $x_\mathrm{proj}$.
  Therefore, for this
effect to be consistent with arising from a feature of nucleon-nucleon collisions, transverse energy production at small
angles should decrease strongly and continuously with increasing $x_\mathrm{proj}$.

  Better understanding of the interplay between soft and hard physics and centrality in \pPb\ can be gained also from the measurement of $Z$~boson production 
\cite{ATLAS:2014cha}. An important component of the measurement is the centrality bias correction \cite{Perepelitsa:2014yta} which should reduce the correlation 
between hard process rates and the magnitude of the soft event activity. After applying the centrality bias correction the \Ncoll\ scaled yield of $Z$~bosons is 
approximately centrality independent, consistent with a production mechanism that scales with the number of binary collisions -- though this scaling is less good for 
Glauber-Gribov Colour Fluctuation model \cite{Guzey:2005tk,Alvioli:2013vk} than for the simple Glauber model.
  The $Z$~boson production cross section measured as a function of rapidity and $x_\mathrm{Pb}$ is systematically larger compared to 
  predictions based on perturbative QCD calculations even after including effects from modifications of nPDF as predicted by EPS09. 

  While the next-to-leading order (NLO) pQCD with EPS09 does not fully describe the rapidity dependence of the $Z$~boson production, it was found to describe well the rapidity dependence of 
the prompt $J/\psi$ production, measured over \pt\ interval of 8-30 GeV \cite{Aad:2015ddl}. In that measurement, the non-prompt and prompt $J/\psi$ production cross-section was evaluated as 
a function of \pt\ and rapidity. Also measured was the $\psi(2)$ \cite{ATLAS:2015pua}. The non-prompt $J/\psi$ and $\psi(2)$ cross-sections were found to be described, 
though with large theoretical uncertainties, by FONLL calculations \cite{Cacciari:2012ny} which do not include nuclear effects. Further, the nuclear modification factor, \RpPb , was evaluated 
based on an interpolation of the cross sections at 2.76, 7 and 8 TeV.
  The \RpPb\ of both prompt and non-prompt $J/\psi$ were measured to be above unity by about 10-30\% showing no significant dependence on \pt\ and rapidity. After applying the 
centrality bias correction, the \RpPb 's do not exhibit dependence on number of participating nucleons. The centrality bias correction also weakens the linear 
correlation between the self-normalized ratios of $J/\psi$, $\psi(2)$, and $Z$~boson production and self-normalized transverse energy measured in the forward calorimeter, 
$E_\mathrm{T}^\mathrm{FCal} / \langle E_\mathrm{T}^\mathrm{FCal} \rangle$.

  While the \RpPb\ of $J/\psi$ shows an enhancement with no significant \pt\ dependence, the previously measured charged particle \RpPb\ exhibited a significant 
increase for $\pt \gtrsim 10-20$~GeV reaching a maximum of 1.4 at $\pt \approx 60$~GeV \cite{ATLAS:2013hza}. This unexpected excess in yields of charged particles seen 
across different rapidity bins contrasts with only little modification seen in the measurement of inclusive jet production discussed previously. Thus, a measurement of 
jet fragmentation was performed \cite{ATLAS:2015mla} to shed light on the origin of this effect.
  In that measurement, the \pPb\ data were compared to a \pp\ reference, constructed by extrapolating the measured fragmentation functions in 2.76 TeV \pp\ collisions to 
5.02 TeV. Using this reference, a ratio of fragmentation functions, 
  $R_{D(z)}=D(z)|_{p\mathrm{Pb}} / D(z)|_{pp}$, 
  was constructed. The measured $R_{D(z)}$ exhibits a $z$-dependent 
excess with a maximal magnitude of approximately 10\% for $0.2 \lesssim z \lesssim 0.8$ in jets with $\pt > 80$~GeV. The $z$ and $\pt$ ranges over which the $R_{D(z)}$ 
distributions are enhanced correspond to the same range in transverse momentum where the inclusive charged particle spectrum in \pPb\ collisions is enhanced.

  Not only high-\pt\ probes measured in \pPb\ collisions brought new interesting physics, but also a region of low \pt\ revealed unexpected strong signals. 
Shortly after the first \pPb\ run a double-ridge structure in long-range pseudorapidity correlations was observed in high-multiplicity \pPb\ collisions 
\cite{Aad:2012gla}. In subsequent studies \cite{Aad:2013fja,Aad:2014lta}, the azimuthal structure of such long-range correlations was Fourier decomposed to obtain the 
harmonics $v_n$ as a function of $\pt$ and event activity. The extracted $v_n$ values for $n = 2$ to 5 decrease with $n$. The $v_2$ and $v_3$ values were found to be 
positive in the measured $\pt$ range. The $v_1$ was also measured as a function of $\pt$ and was observed to change sign around $\pt \approx 1.5-2$~GeV and then 
increase to about 0.1 for $\pt > 4$ GeV. The $v_2 (\pt )$, $v_3 (\pt)$ and $v_4 (\pt)$ were compared to the $v_n$ coefficients in \PbPb\ collisions at 
$\sqrt{s_\mathrm{NN}} = 2.76$~TeV with similar event multiplicities. Reasonable agreement was observed after accounting for the difference in the average $\pt$ of 
particles produced in the two collision systems. This agreement suggests that the long-range ridge correlations in high-multiplicity \pPb\ collisions and peripheral 
\PbPb\ collisions are driven by similar dynamics.

\section{Physics of \PbPb\ collisions}
\label{sec:pbpb}

  Not only \pPb\ collisions but also measurements of high-\pt\ photons and electroweak bosons in \PbPb\ collisions can be used to study 
the nuclear initial state effects and the role of the geometry of colliding nuclei.
  Recently, the prompt photon production \cite{Aad:2015lcb} and the production of $W^{\pm}$ \cite{Aad:2014bha} was measured in \PbPb\ collisions. Earlier, the 
production of $Z$~bosons in \PbPb\ collisions was studied \cite{Aad:2012ew}.

  Inclusive photon yields, scaled by the mean nuclear thickness function, were measured as a function of collision centrality and transverse momentum in two 
pseudorapidity intervals, $|\eta| < 1.37$ and $1.52 \leq |\eta| < 2.37$. The scaled yields in the two pseudorapidity intervals, as well as the ratios of the forward 
yields to those at midrapidity, were compared to the expectations from next-to-leading order perturbative QCD calculations from Jetphox \cite{Catani:2002ny}. The scaled yields agree 
well with the predictions for \pp\ collisions within statistical and systematic uncertainties.
  Both the yields and ratios of prompt photons were also compared to two other pQCD calculations, one which uses the isospin content appropriate to colliding lead nuclei and another 
which includes the EPS09 nuclear modifications to the nucleon parton distribution functions. The data are unable to distinguish between the three 
scenarios. However, the overall consistency of the measured yields with Jetphox expectations for all centrality intervals demonstrates that photon yields in heavy ion 
collisions scale with \Ncoll , that is with the mean nuclear thickness, as expected.

  The $W^{\pm}$~boson production was measured in electron and muon decay channels. The differential production yields and lepton charge asymmetry were each measured as a 
function of the average number of participating nucleons \ANpart\ and absolute pseudorapidity of the charged lepton, $|\eta_l|$, within $|\eta_l|<2.5$. The $W$ boson 
yields scaled by $1/\ANcoll$ were found to be independent of centrality and in a good agreement with NLO pQCD predictions. Due to the different isospin content of \PbPb\ 
compared to \pp\ collisions, the lepton charge asymmetry from $W^{\pm}$ boson decays differs from that measured in \pp\ collisions. The lepton charge asymmetry agrees 
well with theoretical predictions using pQCD at NLO with CT10 \cite{Lai:2010vv} PDF sets with and without EPS09 nuclear corrections. Clearly, further improvements in the experimental 
precision and uncertainties in the theory are needed to establish the existence of nuclear effects. Despite that, these measurements provide further support for the 
interpretation of the clear modification of jet yields in \PbPb\ collisions as a function of centrality, relative to those measured in proton-proton collisions, as 
stemming from energy loss in the hot, dense medium.
 
  First measurement of jet suppression \cite{Aad:2010bu,Aad:2012vca} were followed recently by a precise measurement of jet nuclear modification factor, \Raa , which 
was evaluated as a function of centrality, jet \pt , and jet rapidity \cite{Aad:2014bxa}. 
  The jet yields were measured over the kinematic range of jet transverse momentum $32 < \pt < 500$ GeV, and absolute rapidity $|y| < 2.1$. The jet \Raa\ was found to 
reach a value of approximately 0.5, implying that the jet yields are suppressed by a factor of two in central collisions compared to \pp\ collisions. The \Raa\ shows a 
slight increase with \pt\
 and no significant variation with rapidity. This later observation is in particular striking given the large differences in slopes of jet spectra and differences in 
the jet flavor at different rapidities. Also striking is relatively sizable modification seen in 60-80\% peripheral collisions with the $\Raa \approx 0.8$ for jets 
with $\pt < 100$~GeV.

  A high-precision measurements of jet \Raa\ is only possible with a good understanding of the detector response, in particular the jet energy scale (JES) which was 
elaborated in Ref.~\cite{ATLAS:2015twa}.
  The JES was constructed by calibrating the heavy ion (HI) jets with respect to the standard \pp\ jets using the analysis of \pp\ collision data.
  The JES for the latter is well understood through a combination of $in~situ$ techniques in which the transverse momentum balance between a jet and a reference object 
such as a $Z$ boson or $\gamma$ was measured.
  To establish a JES calibration for the HI jets, \pp\ data at $\sqrt{s} = 8$~TeV recorded in 2012 were analyzed and both \pp\ and HI jets were reconstructed. The 
\pp\ jets were then used as a reference to cross calibrate the HI jets. Applying the HI JES in data from different run periods requires to further consider the 
effects of different detector operating conditions as well as the fact that the flavor fractions are expected to change with beam energy.
  An uncertainty on the absolute JES was then derived for HI jets resulting from the ``baseline'' uncertainty on the reference jets, additional uncertainties in the 
cross calibration procedure itself and the flavor composition and response uncertainties of HI jets. Finally, differences in the response to quenched jets were 
evaluated and a systematic uncertainty which accounts for such differences was derived using the PYQUEN \cite{Lokhtin:2005px} sample. The PYQUEN generator was setup to 
run in two different configurations chosen to match the modifications of fragmentation functions measured in the data.
  The total JES uncertainty in central Pb+Pb collisions was found to be largest at low \pt , approximately 4\% at 30 GeV, and decreasing with \pt\ to values between 2-3\% at 
450 GeV.

  A complementary measurement to the jet \Raa\ is the \Raa\ of charged particles \cite{Aad:2015wga}. In that analysis, the charged particle spectra were measured over a wide transverse 
momentum range, $0.5 < \pt < 150$~GeV, and in eight bins in pseudorapidity covering the range of $|\eta| < 2$. 
  The charged particle \Raa 's show a distinct
\pt\ dependence with a pronounced minimum at about 7 GeV. Above 60 GeV,
\Raa\ is consistent with a plateau at a centrality-dependent value which is approximately 0.55 for the 0-5\% most central collisions.
  The \Raa\ distribution is consistent with flat pseudorapidity dependence over the
whole transverse momentum range in all centrality classes.

  Since inclusive charged particles measured at different transverse momenta may come from jets of different original energies, there is no simple mapping between the 
modified charged particle yields and the modification of the jet internal structure. Thus, our understanding of the jet quenching should strongly profit from the direct 
measurement of jet fragmentation published by ATLAS in Ref.~\cite{Aad:2014wha}. In that measurement, it was found that the central-to-peripheral ratios of fragmentation functions 
of jets with $\pt>100$~GeV show a reduction of fragment yield in central collisions relative to peripheral collisions at intermediate $z$ values, $0.04 < z 
< 0.2$ and an enhancement in fragment yield for $z < 0.04$. A smaller, less significant enhancement was observed at large $z$ in central collisions. Similar 
observations were done also for the distributions of transverse momenta of charged particles reconstructed inside jets. From the analysis of measured distributions it 
was concluded for the 0-10\% most central collisions that
  the increase in the number of particles with $0.02 < z < 0.04$ is less than one particle per jet. A decrease of about 1.5 particles per jet was observed for $0.04 < z 
< 0.2$. Further it was concluded that in the 0-10\% most central collisions a small fraction, $<$2\%, of the jet transverse momentum is carried by the excess particles 
in $0.02 < z < 0.04$ for central collisions, but that the depletion in fragment yield in $0.04 < z < 0.2$ accounts on average for about $14\%$ of jet \pt . 
  These observations should directly help to improve understanding of modifications of parton showers in the medium.

  To improve understanding of the path length dependence of the jet quenching and the role of fluctuations in the jet quenching, three more differential measurements have been 
performed: the measurement of azimuthal dependence of inclusive jet yields \cite{Aad:2013sla} and more recently 
the measurement of correlation between dijet-asymmetry and event-shape variables \cite{ATLAS:2015lla}
and
the measurement of neighbouring jet production \cite{Aad:2015bsa}. The results of these three measurements are described in the next paragraphs.

  In the measurement of the azimuthal dependence of inclusive jet yields, the variation of inclusive jet suppression as a function of relative azimuthal angle, $\Delta 
\phi$, with respect to the elliptic event plane was measured for jets with $\pt > 45$~GeV in six different collision centrality bins. The variation of the jet yield 
with $\Delta \phi$ was characterized by the parameter, $v_2^\mathrm{jet}$, and the ratio of out-of-plane ($\Delta \phi \sim \pi/2$) to in-plane ($\Delta \phi \sim 0$) 
yields. Non-zero $v_2^\mathrm{jet}$ values were measured in all centrality bins for jets with $\pt < 160$~GeV. The jet yields were observed to vary by as much as 20\% 
between in-plane and out-of-plane directions and a significant $v_2^\mathrm{jet}$ of 2-5\% was observed even at very large jet \pt . This $v_2^\mathrm{jet}$ is a direct 
measure of the differential energy loss suffered by hard partons as they traverse different path-lengths of the medium in events characterized by the same collision 
centrality.

  Motivated by $v_2^\mathrm{jet}$ measurements, a similar analysis was performed for the dijet asymmetry. 
  The dijet asymmetry, \aj , was studied as a function of angle between the leading-jet and the elliptic event-plane.
  These measurements effectively study the path-length dependence of \aj\ by requiring the jet pair to traverse different lengths of the medium.
  The dependence of the \meanaj\ on the event-plane angle was quantified by calculating the second Fourier coefficient of the \meanaj\ azimuthal distribution, termed 
$c_2$. The measured $c_2$ signal is quite small ($\leq$2\%), however, it is consistently negative, indicating a slightly larger asymmetry when the dijet pair is 
oriented out-of-plane than in-plane.
  This measurement was further extended by repeating the analysis when constraining the shape of the collision geometry by selecting events based on the magnitude of 
the second-order flow harmonic quantified by the magnitude of the $q_2$ vector.
  Within a given centrality interval, events with large $q_2$, i.e. events with a more elliptic geometry, show an increase in the $c_2$ for the 20-30\% and 30-40\% 
centrality bins. In other centrality intervals, the statistical precision of the results, obtained with additional binning in $q_2$, is insufficient to conclude the 
observation of any systematic dependence of \meanaj\ on the event-plane angle.

  To further constrain the energy loss models, the measurement of the correlations between jets that are separated by small relative angles was performed.
  The measured neighbouring jet pairs result primarily from hard radiation by the parton that occurs early in the process of the shower formation. Generally, two 
neighbouring jets originating from the same hard scattering should have more similar path lengths in the medium compared to the two jets in the previous dijet 
measurement. Therefore measuring neighbouring jets could probe differences in their quenching that do not result primarily from difference in path length as well as put 
some constraints to models in which a part of the parton shower radiates coherently in the response to the medium.
  The production of pairs of correlated jets was quantified using the rate of neighbouring jets that accompany a test jet, \RdR . This observable was used in past by 
D$\emptyset$ collaboration to measure the strong coupling constant, $\alpha_s$, and to test its running over a large range of momentum transfers \cite{Abazov:2012lua}.
  The \RdR\ was evaluated both as a function of test-jet \Et\ and neighbouring-jet \Et\ . 
  A significant dependence of \RdR\ on collision centrality was observed in both cases, suggesting a suppression of neighbouring jets which increases with increasing 
centrality of the collision. The centrality dependence of the suppression was further quantified using the central-to-peripheral ratio of \RdR\ distributions, \rhoRdR . 
The trends seen in \rhoRdR\ evaluated as a function of neighbouring-jet \Et\ indicate a decrease in suppression with increasing neighbouring-jet \Et . The \rhoRdR\ 
evaluated as a function of test-jet \Et\ exhibits a suppression reaching values of 0.5-0.7 in 0-10\% central collisions and does not show any strong dependence on \Et .

  The weak $c_2$ signal in the dijet asymmetry, the significant single jet $v_2$ and characteristic behavior seen in the neighbouring jet production can provide strong 
constraints in particular on the path length dependence of the energy loss.

  The azimuthal anisotropy of particle emission is generally a useful tool to study the properties of the hot and dense matter created in heavy ion collisions. At high 
transverse momenta this anisotropy is understood to result from the path-length dependent energy loss of jets as they traverse the deconfined matter as discussed in previous 
paragraphs. 
  At low transverse momenta ($\pt \lesssim 3-4$~GeV), this anisotropy results from a pressure driven anisotropic expansion of the created matter, with more particles 
emitted in the direction of the largest pressure gradient. 
  In Pb+Pb collisions, ATLAS has performed event-averaged measurement of elliptic flow and higher order flow harmonics \cite{ATLAS:2011ah,ATLAS:2012at,Aad:2014eoa,Aad:2014vba},
  the event-by-event flow measurements \cite{Aad:2013xma}
  and the measurement of event plane correlations \cite{Aad:2014fla}.
  Recently, ATLAS has performed also the measurement 
  of correlations between different flow harmonics in \PbPb\
collisions \cite{Aad:2015lwa}.
  In that measurement, it was found that $v_3$ is anticorrelated with $v_2$ and this anticorrelation is consistent with similar anticorrelations between the 
corresponding eccentricities $\epsilon_2$ and $\epsilon_3$. On the other hand, it was observed that $v_4$ increases strongly with $v_2$, and $v_5$ increases strongly 
with both $v_2$ and $v_3$. The trend and strength of the $v_m - v_n$ correlations for $n = 4$ and 5 were found to disagree with $m - n$ correlations predicted by 
initial-geometry models. Instead, these correlations were found to be consistent with the combined effects of a linear contribution to $v_n$ and a nonlinear contribution that is 
a function of $v_2$ or of $v_2 v_3$, as predicted by hydrodynamic models.
  
  A complementary measurement to the measurement of correlations of azimuthal harmonics is the measurement of two particle pseudorapidity correlations 
\cite{ATLAS:2015kla}.
  Two particle pseudorapidity correlations can probe e.g. the rapidity profile of the initial-state fireball. This ATLAS measurement was done in the pseudorapidity 
range of $|\eta|< 2.4$ with charged particles having $\pt > 0.5$~GeV.
  The measured two particle correlation functions show a broad ``ridge-like'' shape along $\eta_1 = \eta_2$ and a depletion at around $\eta_1 = -\eta_2$.
  The correlation functions were expressed in terms of $\eta_- \equiv \eta_1 - \eta_2$ and $\eta_+ \equiv \eta_1 + \eta_2$ which allows to separate the short-range 
correlation effects (centered around $\eta_- \approx 0$) from the genuine long-range correlations.
  Further, the measured two particle correlation functions were expanded in terms of products of Legendre polynomials, $T_n(\eta_1)T_m(\eta_2)$, and corresponding 
coefficients, $a_{n,m}$, were extracted. The first term, $\sqrt{ \langle a_n^2 \rangle}$, is connected with the asymmetry between forward and backward particle 
production which was confirmed to be large as seen previously in other measurements. The second term reflects the even-by-event fluctuation of the width of charged 
particle multiplicities, $N(\eta)$.
  Significant mixed coefficients $\langle a_n, a_m \rangle$ are also observed. The most significant group of mixed coefficients $\langle a_n, a_{n+2} \rangle$ is found 
to be negative, implying that $a_n$ and $a_{n+2}$ are generally anticorrelated. The magnitudes of $\langle a_n, a_{n+2} \rangle$, were found to decrease quickly for 
larger $n$.
  These extracted coefficients are generally found to increase for peripheral collisions, which is consistent with the increase of the multiplicity fluctuation for 
smaller collision systems.

  The above presented correlation measurements should help to better understand the (fluctuating) initial conditions of expanding medium and to further test the 
hydrodynamical paradigm.

  The reader is welcome to follow the details in full publications which are listed in \cite{PublicResults}.

\section*{Acknowledgment} 

  This work was supported by Charles University in Prague, projects PRVOUK P45 and UNCE 204020/2012, and by MSMT CR (project INGO LG13009).

\section*{References}

\bibliographystyle{elsarticle-num}
\bibliography{MartinSpousta_forArxiv}

\end{document}